# Deep learning analysis of polaritonic waves images

Suheng Xu[1], Alexander. S. McLeod[1], Xinzhong Chen[2], Daniel J. Rizzo[1], Bjarke S. Jessen[1,3], Ziheng Yao[2], Zhicai Wang[2], Zhiyuan Sun[1], Sara Shabani[1], Abhay N. Pasupathy[1], Andrew J. Millis[1,4], Cory R. Dean[1], James C. Hone[3], Mengkun Liu[2,5], D. N. Basov[1]

*[1]Department of Physics, Columbia University, New York, New York 10027, USA*

*[2]Department of Physics and Astronomy, Stony Brook University, Stony Brook, New York 11794, USA*

*[3]Department of Mechanical Engineering, Columbia University, New York, New York, 10027, United States*

*[4]Center for Computational Quantum Physics, Flatiron Institute, New York, New York 10010, USA*

*[5]National Synchrotron Light Source II, Brookhaven National Laboratory, Upton, New York 11973, USA*

## Abstract

**Deep learning (DL) is an emerging analysis tool across sciences and engineering. Encouraged by the successes of DL in revealing quantitative trends in massive imaging data, we applied this approach to nano-scale deeply sub-diffractional images of propagating polaritonic waves in complex materials. We developed a practical protocol for the rapid regression of images that quantifies the wavelength and the quality factor of polaritonic waves utilizing the convolutional neural network (CNN). Using simulated near-field images as training data, the CNN can be made to simultaneously extract polaritonic characteristics and materials parameters in a timescale that is at least three orders of magnitude faster than common fitting/processing procedures. The CNN-based analysis was validated by examining the experimental near-field images of charge-transfer plasmon polaritons at Graphene/α-$RuCl_3$ interfaces. Our work provides a general framework for extracting quantitative information from images generated with a variety of scanning probe methods.**

## Introduction

Polaritons, hybrid collective modes of light and polar resonances in materials, are of high interest to the field of nanophotonics and quantum information processing.[1–4] Various types of polaritons have been observed in van der Waals materials, where photons couple to plasmons, phonons, excitons, or Cooper pairs and form robust hybrid modes.[5–14] The wavelength of polaritons $\lambda_p$ is reduced compared to the wavelength of free space photons $\lambda_0$ by orders of magnitude.[1] It is therefore said that polaritons are highly confined on sub-wavelength length scales. An immense utility of polaritons is that the response function of a material can be inferred from the visualization of the polaritonic interference patterns.[7,15–17] A common way to launch and image polaritons is based on the scattering-type scanning near-field optical microscope (s-SNOM).[4,18] Here, the tip of an atomic force microscope (AFM) scatters the free propagating light into evanescent waves with high in-plane momentum on the order of the inverse of tip radius, which simultaneously excites the polaritons and outcouples confined near-field into the far-field (shown schematically in Figure 1(a)). Raster scanning the illuminated AFM tip near a material surface thus generates a map of the local near-field signal with a high spatial resolution (~10 nm). Here, polaritons present as periodic fringes in the nearfield amplitudes. These images can be used to reconstruct the energy-momentum dispersion by fitting the frequency-dependent spatial line profiles with a functionals ansatz.[7,16,17] However, this fitting procedure is time-consuming and is difficult to automate. The predicament stems from the complex interferences of polaritons, especially when a small sample size in combination with low damping permit multiple cavity-like reflections by polaritonic wave.[5,6,11,15,19–21] In this work, we have developed a deep learning (DL) routine for the fast extraction of crucial polaritonic parameters from nano-IR images with little human assistance. Furthermore, we document the utility of this approach by analyzing plasmonic fringes at the interface to two atomically layered van der Waals materials: graphene and $\alpha$-$RuCl_3$.

Deep learning is a branch of artificial intelligence that is commonly employed for rapid quantitative analysis across disciplines.[22–24] Starting in the 1990s with experimental particle physics, DL has impacted many sub-fields of physics.[25–30] Notably, DL algorithms reveal new data trends with minimal assumptions[31] and enabled advances in scanning probe microscopy. This includes the ability to identify new translational-symmetry-breaking states[32], quantitative analysis of the optical signal in s-SNOM experiments[33,34], accelerated analysis of raster-scanned images[35], along with automated scanning probe microscopy without user control.[36] Here, we analyzed the polariton images with a convolutional neural network (CNN) that is commonly used for image classification.[37,38] We demonstrated that CNN can capture key features in nano-IR images of polariton interference patterns. The CNN trained with only synthetic data was nevertheless capable of accurately extracting wavelengths and quality factors from the experimental nano-IR images. We focus on the analysis of charge transferred plasmon polaritons (CPPs) in graphene/$\alpha$-$RuCl_3$ as a case study.[21] We also demonstrate that the CNN is capable of dealing with more complex polaritonic interference patterns, that are formed by the superpositions of modes with different wavelengths. Our results indicate that DL-based analysis of polariton images offers a highly efficient and high-fidelity alternative to traditional methods, providing the conceptual bedrock for similar future applications.

## Scanning polariton interferometry

In this section, we provide a primer of nano-infrared visualization of polaritonic waves. Figure 1(a) depicts the working principle of scanning polariton interferometry based on s-SNOM. The tip-launched surface polaritons (SPs) can be reflected by edges of the sample, defects, or grain boundaries, and other forms of structural/electronic inhomogeneities.[19] The reflected SPs interfere with the outgoing SP wave, forming a standing wave pattern. Since the far-field signal is determined by the local electric field strength under the AFM tip, raster scanning the tip leads to a spatial interferogram. Both the amplitude and phase of the tip-scattered signal can be obtained in the pseudo-heterodyne detection scheme.[18,39]. In s-SNOM, the AFM tip oscillates at the mechanical resonance frequency Ω and the background signal can be highly suppressed by demodulating the scattered signal to higher harmonic orders of the tapping frequency ($n\Omega$, $n \geqslant 2$). In Figure 1(b), we present the near-field amplitude measured for an hBN/Graphene/α-$RuCl_3$/$SiO_2$ heterostructure near the graphene edge. Here, the periodic fringe pattern emanating perpendicular to the edge corresponds to the tip-launched SPs.[21]

The propagating wave nature of polaritons can be characterized by a complex wavevector $q_p = q_1 + iq_2$, which depends on the effective dielectric function of the environment, in-plane conductivity, and frequency.[1,6] The wavelength of the polaritons is determined by the real part of the wavevector $\lambda_p = \frac{2\pi}{q_1}$ and the quality factor is defined by the ratio between the real part and the imaginary part of the wavevector $Q = \frac{q_1}{q_2}$. Since s-SNOM visualizes the standing wave pattern of polaritons, the distance between two fringes is half of $\lambda_p$ for tip-launched round-trip polaritons. By evaluating the wavelength and quality factor from near-field intensity profiles, the local electronic properties (e.g., Fermi level and scattering rate) can be obtained.[6,11]

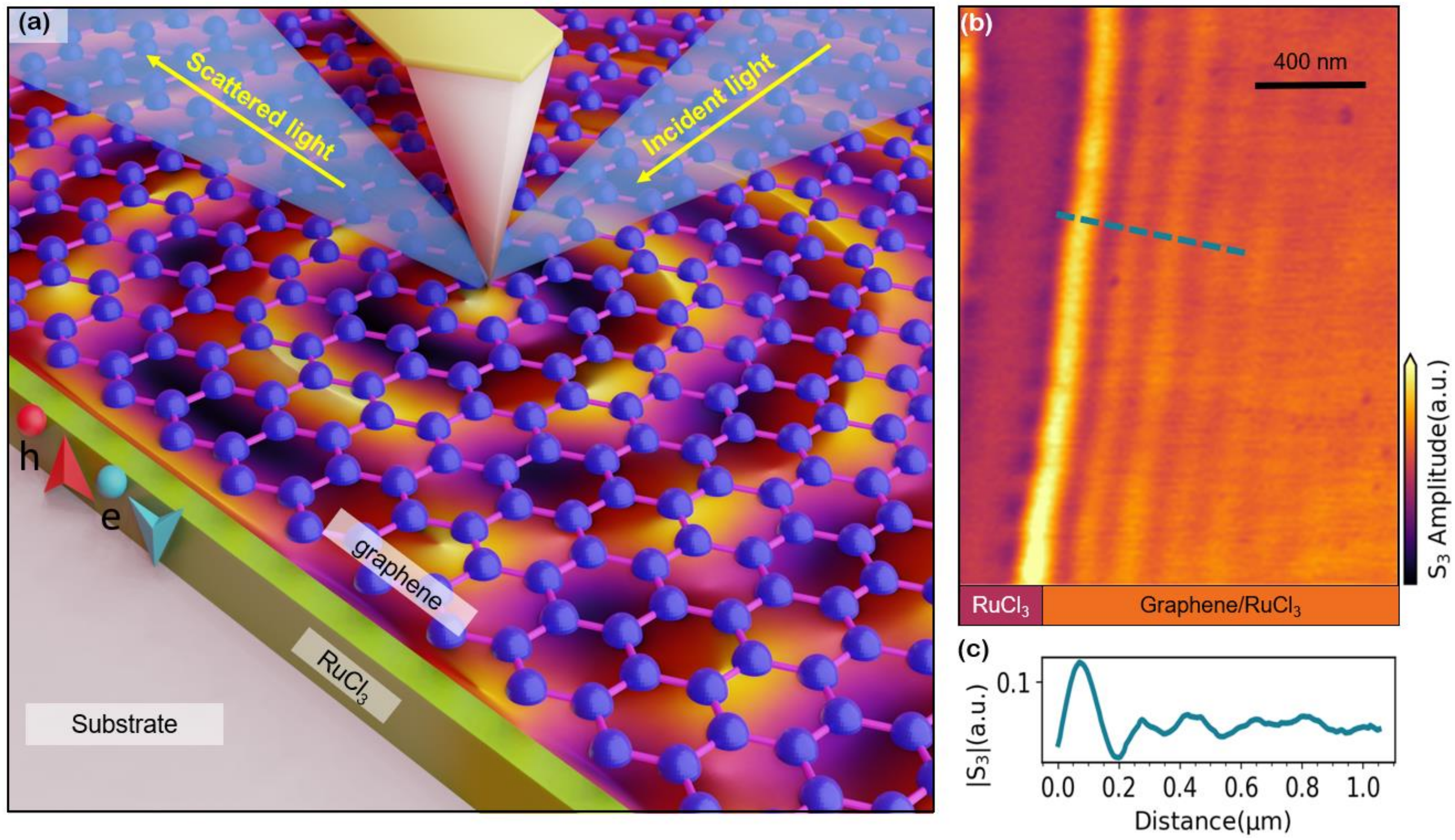

**Figure 1. Scanning polariton interferometry of an hBN/Graphene/α-$RuCl_3$/$SiO_2$ heterostructure. (a) Schematics of surface polaritons excited by focused continuous wave (CW) light on an AFM tip. The local near-field response under the tip is measured by collecting the scattered field in the far-field. (b) Map of the near-field amplitude on hBN/Graphene/α-$RuCl_3$/$SiO_2$ heterostructure (T=80 K, $\omega = 898\ \mathrm{cm}^{-1}$). The profile of the oscillating fringes in the bulk indicates the presence of surface plasmon polaritons. (c) Line profile of surface plasmon polaritons along the dashed line in (b).**

## Deep learning analysis of charge-transfer plasmon polaritons

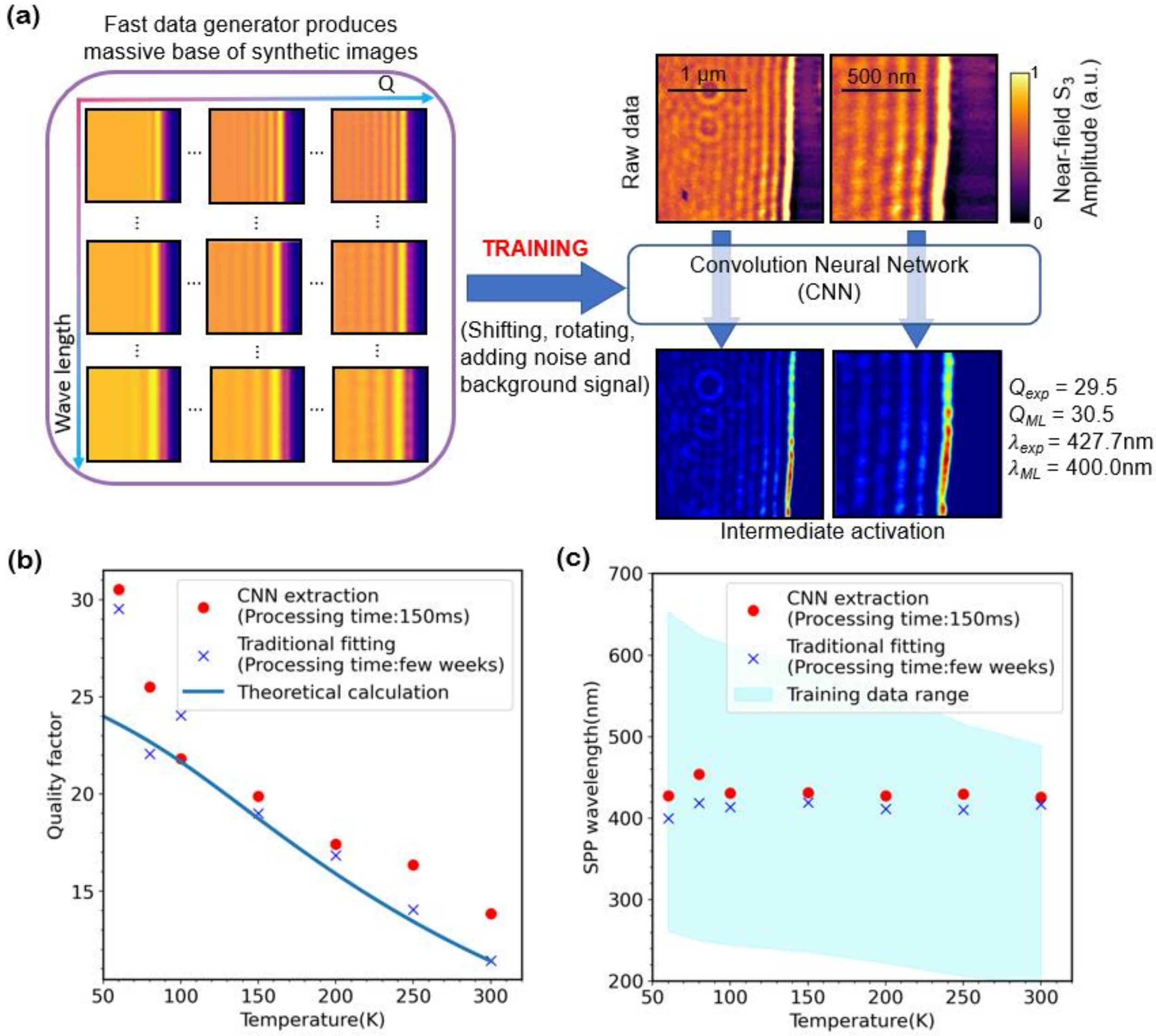

**Figure 2. Training, application, and extraction results of CNN-based analysis of polaritonic images. (a) Workflow of CNN training and applications. A numerical solver package is used to generate a massive set of polaritonic images labeled with different combinations of wavelengths and quality factors. After data augmentation and adding noise to the images, the data is fed into the CNN for training. With the trained CNN, both the quality factor and the wavelength are extracted with the experimental data as input. The intermediate activation visualization shows that the CNN captures key features of the polaritonic patterns. (b) Quality factor versus sample temperature evaluated by theoretical prediction (blue solid curve), fitting (blue cross), and CNN extraction (red dot). (c) Wavelength versus temperature evaluated by fitting (blue cross) and CNN extraction (red dot).**

Here, we introduce a CNN-empowered method for analyzing polaritonic images. Fed with 1725 synthetic near-field images generated by numerical simulations, a trained network was capable of extracting both the quality factor and the wavelength from an experimental near-field image within 50 ms. In Figure 2(a), we show the workflow for training and applying CNN for polaritonic image analysis. To generate synthetic near-field images as training data, we employed the recently reported simulation model which successfully reproduced the near-field signal of surface plasmon polariton reflected at a point defects.[21] In the simulation, the near-field antenna probe was approximated as a point dipole. The scattering signal was evaluated by the generalized reflection coefficient of the probe with spatial dependence (see supplemental materials). The area for simulation was on the length scale of 4×4 μm with a spatial resolution of 10 nm. A square plasmonic media was located at the center of the substrate and the near-field intensity was recorded at the edge of the sample. All simulated near-field signals were normalized into a range of 0 to 1 to simplify the network training. Synthetic near-field images were labeled with wavelengths and quality factors varied from 73 nm to 167 nm and from 5 to 30, respectively. The specific range of simulation parameters was chosen to avoid extra interferences due to the polariton reflected by other edges of the sample. Additional background signal and noise were added to the near-field signal on both the sample and the substrate area to mimic the similar situations in near-field experiments (see supplemental material). To endow the CNN with the ability to make accurate extractions regardless of the orientation of sample boundaries in the images, we applied data augmentation to the training dataset. By shifting and rotating the images randomly, we generated near-field images much more similar to experimental results and consequently enhanced the generalization of the CNN. (see supplemental materials). A validation dataset generated with randomly distributed parameters acted as a metric to evaluate the generalization error of the CNN.

As a proof-of-principle demonstration, we applied the trained CNN to the temperature-dependent data of charge transfer plasmon polaritons that were recently observed for α-$RuCl_3$/graphene interfaces.[21] Due to the large work function difference between α-$RuCl_3$ and graphene (6.1 eV versus 4.6 eV, respectively), in excess of $10^{13}$ $cm^{-2}$ charge carriers are transferred across the 2D interface. The net effect of charge transfer is that the Fermi energy in graphene is subsequently shifted by 0.6 eV with respect to the charge neutrality point. The undoped graphene became highly metallic without chemical doping or back-gating, as indicated by the distinct polaritonic interference patterns observed by s-SNOM (Fig. 1(b)). Furthermore, this massive charge transfer may also lead to the emergence of the topological quantum spin liquid state in α-$RuCl_3$.[40,41] Here, we selected square regions in the experimental nano-IR images which included the sample edge and the polaritonic fringe patterns (see supplemental materials). In each nano-IR image, the near-field amplitude is normalized into the range of 0 to 1 before extraction. Since the scales of pixels in the measurement depend on the temperature, the range of the training dataset varies at different temperatures. The extracted quality factors and wavelengths shown in Figure 2(b) and (c) agree well with the values obtained from the traditional fitting method. The small discrepancy between the parameters generated with the two approaches ($<13.5\%$) might come from the intrinsic uncertainty of the CNN, the bias between simulation data and experimental data, and the inhomogeneous nature of samples. The CNN successfully extracted the temperature-dependent quality factors and wavelengths from the raw data with relatively high accuracy. In addition, the overall extraction process took up less than 150 ms for 14 different images, which was of great advantage and efficiency compared with the traditional fitting.

### Convolution neural network for polaritonic near-field image analysis

The superiority of the CNN for polaritonic nano-IR image analysis is attributed to the working principles and the architecture of the CNN.[31] Previously, CNN-based architectures have shown great performance on complex image recognition tasks and have even outperformed human beings in tasks such as facial recognition.[42] As a subtype of the neural network, the matrix operations in several layers are replaced by convolutional operations. The typical architecture of the CNN is presented in Figure 3(a). An input near-field image first goes through a series of convolution layers and pooling layers, which extract the local features in the image and decrease its dimensions. The information in the original data is distilled into more abstract patterns in deeper convolution layers. After going through the last convolution layer, the tensors are flattened and onward to fully connected dense layers whose outcomes are the quality factor and wavelength. Compared with the fully connected dense layers, the convolution and pooling operations in the CNN improve the speed and accuracy of the CNN. Since the convolution kernel can effectively capture the local patterns, and operate several convolution layers simultaneously, the CNN will eventually gain a powerful feature extraction capability of the global information. In addition, fewer free parameters are required to train the CNN compared with the fully connected neural network. Though the degrees of freedom of the network are decreased, the network maintains the same performance for extracting polaritonic parameters. Moreover, the CNN learns the pattern in a translationally invariant way and the extraction result is robust to small translations of the input features.[43] Hence, the near-field images do not need to be manually rotated or positioned as a part of the parameter extraction procedure.

The details of our CNN architecture are presented in the supplementary information. (See supplemental materials) The capacity of the network could be altered provided the complexity of the training dataset increases. We chose the rectified linear unit (ReLU) as the activation function, which introduced the nonlinearity between the input and the output of the neurons. The loss function was defined as the Mean Absolute Error (MAE) between the extracted values and the labels. The Root Mean Square Prop algorithm was applied to optimize the network in the training process.[44] In each epoch, all the training data were fed into the CNN and the free parameters in the CNN are tuned to reduce the loss function based on the chosen algorithm. The network training process converges after sufficient epochs and the signature of convergences could be visualized by the MAE loss as a function of epochs which is shown in Fig 3(b). The total training time was around 15 hours. The network training was boosted based on the GPU acceleration (GTX 1050Ti) and the total time for training can be further shrunk with the assistance of more powerful computational hardware. To provide a reference for future CNN development, we also studied the relationship among the MAE loss, training time, and the training data volume. The latter results are summarized in Fig 3(c). The MAE loss evaluated by the same set of validation datasets monotonically decreased when the training dataset was larger while the training time was still in a reasonable range of 15 hours and the number of data points was 3500. However, since the validation dataset was also generated by the same simulation package, the MAE can not fully evaluate the accuracy of the CNN to make extraction on realistic data. To unambiguously evaluate the accuracy of a CNN, a validation dataset composed of labeled experimental data should be applied in the future.

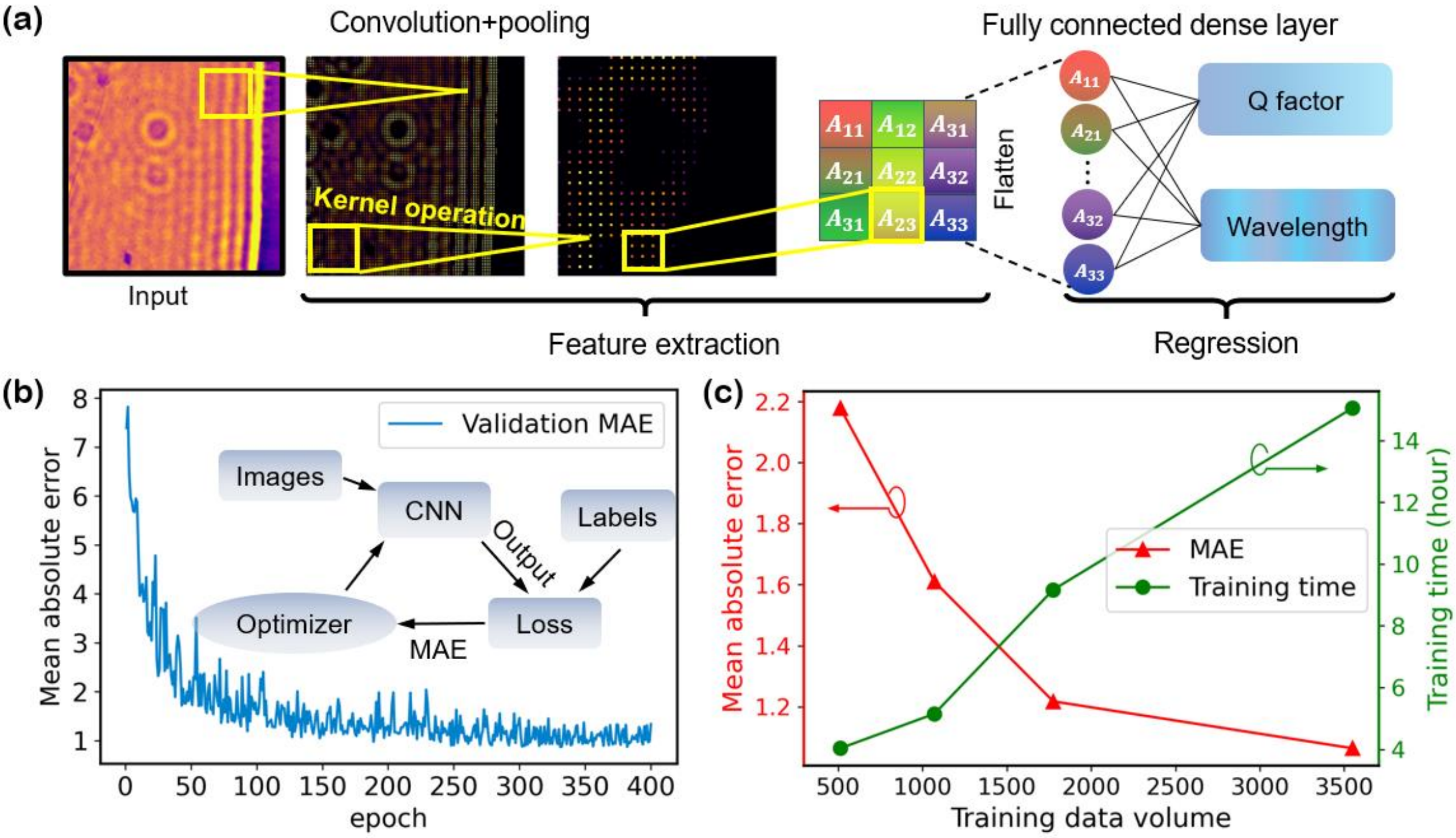

**Figure 3. (a) Schematics of CNN architecture for polaritonic near-field image analysis. The CNN is composed of two parts, which aim at feature extraction and numerical regression, respectively. The Convolution and max-pooling operations extract and abstract the local patterns in the image and the output tensor is flattened into an array, which becomes the input for the following fully connected dense layer for regression. The outputs of the CNN are the quality factor and the wavelength of the polariton in the near-field image. Here, the convolution operations are visualized by the intermediate activation visualizations (see supplementary materials). (b) The mean absolute error as a function of epoch in the training process. Inset: A simplified version of the flowchart for the network training. (c) The mean absolute error and training time as the function of training data volume.**

**CNN for multi-wavelength data extraction**

The CNN-based parameter extraction can be generalized to different types of data, even if the data become much more complex and formidable for traditional fitting methods. In this section, we show that CNN is capable of quantifying polaritonic oscillations with multiple wavelengths simultaneously present in the same field of view of nano-IR images. For example, complex profiles of near-field amplitude composed of multiple waveguide modes are commonly observed in hyperbolic vdW materials.[8,16,45] When the thickness of the crystal is relatively large, the air mode and multiple waveguide modes can be simultaneously excited and probed by s-SNOM.[46,47] However, the analysis tools suitable for processing the multi-wavelength data are limited. Fourier transformation (FT) is widely applied to separate multiple modes in momentum space[7,9,47–49] but its accuracy suffers when the polaritonic data are impacted by losses with only few fringes seen in raw images.

The Schematics of the multiple wavelengths extraction based on CNN are shown in Fig. 4(a). We created the multi-wavelength dataset by superposing two single-wavelength polariton images. Each composite image was labeled with two wavelengths corresponding to the two individual polariton images. The ranges of labels were evenly distributed from 260 nm to 400 nm and from 460 nm to 600 nm for $\lambda_1$ and $\lambda_2$, respectively. The quality factor for each polaritonic near-field image was randomly distributed from 10 to 15. Noise and signal on the substrate were generated in a similar manner and were shown in the Supplemental materials. The CNN was then trained with 900 near-field images and the MAE went below 5 nm by the end of the training process.

In Fig. 4(a), we also show the generation and extraction processes for one multi-wavelength near-field image. The extracted wavelengths are consistent with the labels of synthetic images. Moreover, we also observed from the intermediate activation visualization that the CNN was able to extract at least two fringes corresponding to each wavelength in the feature extraction process. To further present the advantage of the CNN for multi-wavelength extraction, we generated 10 synthetic near-field images that are shown in Figure 4(b). The parameters for the simulation are selected along the two curves presented in Figure 4(e). These 10 synthetic images are not within the training dataset due to the random noise and different parameters that don't present in the training dataset. In Figure 4(c), we showed the extracted line profiles and the multiple modes corresponding to different wavelengths are hard to be directly recognized by human eyes. The Fourier transform (FT) of the line scan (Figure 4(d) showed two distinct peaks and the momenta for different modes can be extracted by recording the position of peaks in the $q$ axis. The comparison between the FT and the CNN extraction results is summarized in Figure 4(e). The CNN demonstrates an overall consistency while the FT deviates from the ground truth especially in the low momentum region. Again, the CNN-based extraction for multi-wavelength polariton images doesn't require any image preprocessing but enables direct analysis on raw nano-IR images in the time scale of milliseconds. To sum up, we have presented the unprecedented accuracy and efficiency for multi-wavelength extraction based on CNN. The preliminary results imply that CNN is an important candidate to analyze polariton images with high complexity.

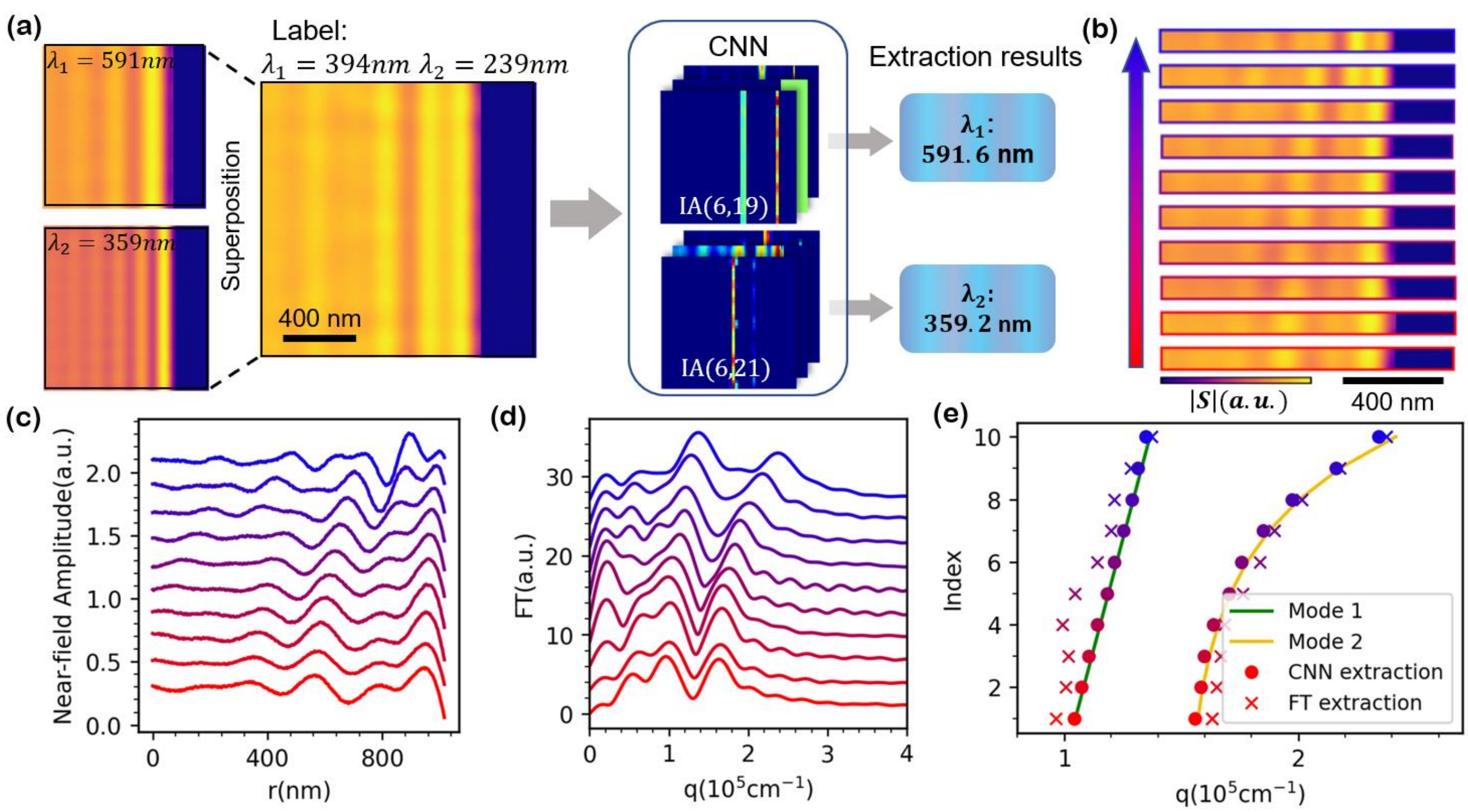

**Figure 4. CNN-based multi-wavelength parameter extraction (a) Schematics of the CNN-based multi-wavelength parameter extraction from nano-IR images. The multi-wavelength data were generated by superposing two simulated single-wavelength polaritonic images and were labeled with the two wavelengths. A few outputs from intermediate activations (IA channel 19 and 21 in layer 6) showed that the CNN extracted the fringes correspond to individual polariton modes. (b) Synthetic polaritonic images generated by combinations of two different wavelengths. (c) Line profiles of the synthetic nano-IR images shown in (b). (d) Fourier Transform (FT) of the line profiles in (c). (e) Comparison among the ground truth, CNN extraction, and FT extraction results.**

## Conclusions and Outlook

In this study, we demonstrated that the CNN can efficiently extract the key quantitative parameters from polariton nano-IR images including wavelengths and quality factors. The current CNN trained with synthetic data successfully extracted the temperature-dependent quality factors and wavelengths for the charge transfer plasmon polariton at the graphene/α-$RuCl_3$ interface. Furthermore, we also showed the CNN can potentially extract multiple wavelengths that correspond to different waveguide modes from more complex interference patterns. The CNN-based polaritonic near-field analysis is able to extract the wavelengths and quality factors over milliseconds, which significantly accelerates the data processing. The large improvement in terms of parameter extraction is fundamentally important for future scanning polariton interferometry studies where a large amount of data is expected to be generated on a routine basis.

The current CNN is trained to work with the interference patterns of polariton that are scattered by straight edges. An extension to circular polaritonic patterns caused by point defects or curved boundaries is rather trivial. Furthermore, inactive channels in the convolution layers are well suited to adapt the current CNN recognizing even more complex interference patterns (see supplemental materials). Moreover, errors in the CNN-based procedure can be reduced by including real data in the training set. (known as Transfer Learning)[50] In addition, more sophisticated DL models and algorithms (e.g. Resnet[51] and Senet[52]) can be applied to achieve more ambitious goals by increasing the number of convolutional layers and introducing the channels' interactions.

In this work, we limited ourselves to the analysis of nano-IR polariton images, while this CNN-based regression analysis approach is generic to parameter extraction tasks for other nanoprobe measurements. In a recent nanoprobe study, the tomography, local conductivity, and local magnetism can be simultaneously resolved by utilizing the multi-messenger nanoprobes.[53] The multi-messenger measurements enlarge the dimensionality of the data and might lead to more accurate and informative data mining based on DL. Moreover, the nano-spectroscopic data also possesses copious information about tomography[54,55], phase transitions[56,57], and local electronic structures of the materials[58,59]. Besides, the DL-based image analysis can be extended to different scanning probe measurements such as scanning tunneling microscopy, where the quasiparticle interference manifests itself as periodic spatial variations in the local density of states.[60–62] The DL-based analysis of these high-dimensional nano-spectroscopic data is a promising direction for future investigation.

**Acknowledgments**

Research at Columbia on graphene/RuCl3 interfaces is supported at Columbia was supported by the US Department of Energy (DOE), Office of Science, Basic Energy Sciences (BES), under award no. DE-SC0018426. The development of new nanofabrication and characterization techniques enabling this work has been supported by the US DOE Office of Science, BES, under award DE-SC0019300. The development of the universal cryogenic platform used for scanning probe measurements is supported as part of the Energy Frontier Research Center on Programmable Quantum Materials funded by the US Department of Energy (DOE), Office of Science, Basic Energy Sciences (BES), under award no. DE-SC0019443. The development of ML protocols at Columbia, Stony Brook, and Brookhaven is supported by the U.S. Department of Energy, Office of Science, National Quantum Information Science Research Centers, Co-design Center for Quantum Advantage (C2QA) under contract number DE-SC0012704. Mengkun Liu was partially supported by the RISE2 node of NASA's Solar System Exploration Research Virtual Institute under NASA Cooperative Agreement 80NSSC19MO2015.

## Supplemental material for

## Deep learning analysis of polaritonic waves images

### Simulation method

To generate predicted real-space images shown in the main text, we apply a semi-analytic method that approximates the near-field scattering signal from a 2D material as proportionate (to first-order) by the *z*-polarization of a polarizable dipole raster-scanned (at a height $z = z_{\rm dp}$) some tens of nanometers over the surface of a sample (at $z = 0$):

$$S(\boldsymbol{\rho}_{\rm dp}) \sim p_z \approx \alpha E_{{\rm ref},z}\,(\boldsymbol{\rho}_{\rm dp}, z = z_{\rm dp}). \qquad \text{(S1)}$$

Here $\alpha$ denotes the dipole polarizability, $E_{{\rm ref},z}$ denotes the z-component of the electric near-field reflected by the sample in response to the incident dipole field, and $\boldsymbol{\rho}_{\rm dp}$ denotes evaluation at the in-plane coordinate of the probe. Although this expression represents only the first term in a Born expansion of the full self-consistent dipole polarization[1] , a similar conceptual treatment was previously shown to faithfully replicate the polaritonic near-field response of two-dimensional materials as measured by scanning near-field optical microscopy[2] . Here we summarize the key points enabling our calculation of Eq. S1 in the quasi-electrostatic approximation, and defer more detailed discussion to forthcoming work.

We recast Eq. S1 in a form reminiscent of the local photonic density of states[3] measured at the location $\boldsymbol{r}_{\rm dp} = (\boldsymbol{\rho}_{\rm dp}, z_{\rm dp})$ of our dipole probe:

$$S(\boldsymbol{\rho}_{\rm dp}) \sim \int_{z>0} dV\ \hat{\jmath}_{\rm dp} \cdot \vec{E}_{\rm ref} = \int_{z>0} dV\ \nabla \cdot \hat{\jmath}_{\rm dp}\,\Phi_{\rm ref} \propto \int_{z>0} dV\ -\varrho_{\rm dp} \cdot \Phi_{\rm ref} \qquad \text{(S2)}$$

Here $\hat{\jmath}_{\rm dp}$ denotes the unit vector oriented along the direction of the point dipole current, $\varrho_{\rm dp}$ denotes the instantaneous charge distribution associated with the dipole, and $\Phi_{\rm ref}$ is the electrostatic potential for the reflected field given by $\vec{E}_{\rm ref} = -\nabla\Phi_{\rm ref}$. Now $S(\boldsymbol{\rho}_{\rm dp})$ can be evaluated entirely in the plane $z = 0$ by identifying the "incident" electrostatic potential generated by the dipole through $\varrho_{\rm dp} = -\frac{1}{4\pi}\nabla^2\Phi_{\rm dp}$ and integrating Eq. S2 by parts, yielding:

$$\int_{z>0} dV\ -\varrho_{\rm dp} \cdot \Phi_{\rm ref} = \frac{1}{4\pi}\left[\int_{z=0^+} dA\ (-\hat{z} \cdot \nabla\Phi_{\rm dp}\,)\Phi_{\rm ref} - \int_{z>0} dV\,\nabla\Phi_{\rm dp} \cdot \nabla\Phi_{\rm ref}\right]$$

$$= \frac{1}{4\pi}\int_{z=0^+} dA\ \left(\Phi_{\rm dp}\partial_z\Phi_{\rm ref} - \partial_z\Phi_{\rm dp}\Phi_{\rm ref}\right). \qquad \text{(S3)}$$

Here we have applied the source-free condition $\nabla^2\Phi_{\rm ref} = 0$ in the volume $z > 0$. Eq. (S3) represents an approximation for the signal $S(\boldsymbol{\rho}_{\rm dp})$ when $\Phi_{\rm dp}$ is produced from a dipole-like probe at $\boldsymbol{r}_{\rm dp}$. Further simplification is admitted by the fact that $\Phi_{\rm ref} = -\hat{R}\Phi_{\rm dp} \equiv -\Phi_{\rm R}$, with $\hat{R}$ a generalized reflection operator. Moreover, for scalar potentials $\Phi_{1,2}$ harmonic (*viz.* source-free) in the plane of integration, $\int dA\,\Phi_1\partial_z\Phi_2 = \pm\int d^2q\ |\boldsymbol{q}|\ \tilde{\Phi}_1\tilde{\Phi}_2$, where tilde quantities represent in-plane Fourier transforms with respect to the momentum $\boldsymbol{q}$, and $\pm$ correspond to the cases where $\Phi_2$ is sourced from $z > 0$ or $z < 0$, respectively. With these considerations, Eq. (S3) reduces to:

$$S(\boldsymbol{\rho}_{\rm dp}) \sim \frac{1}{2\pi}\int d^2q\ |\boldsymbol{q}|\ \tilde{\Phi}_{\rm dp}\tilde{\Phi}_R = \frac{1}{2\pi}\int dA\,(q * \Phi_{\rm dp})\hat{R}\Phi_{\rm dp} \qquad \text{(S4)}$$

where $(q * \Phi_{\rm dp})$ represents the incident scalar potential spatially convolved at $z = 0^+$ with a sharpening function with Fourier kernel $|\boldsymbol{q}|$. Eq. (S4) represents a norm of the function $\Phi_{\rm dp}$ in the plane $z = 0$ with respect to the composite reflection operator $q * \hat{R}$.

By way of demonstration, we can consider cases where the reflected field is given by $\widetilde{\Phi}_R = r_p(q)\widetilde{\Phi}_{\rm dp}(q)$, with $r_p$ the momentum-resolved Fresnel coefficient for *e.g.* a layered medium with in-plane translational invariance. Applying the in-plane Fourier transform of the dipole potential $\widetilde{\Phi}_{\rm dp}(\boldsymbol{q}) = {\rm e}^{-qz_{dp}}$ at $z = 0$, for such cases Eq. S4 evaluates to $S \propto \int dq\, r_p(q)\, q^2 e^{-2qz_{dp}}$. This is indeed the first-order term in a Born series expansion of the point dipole model widely used to predict near-field observables in the case of multilayered systems[4,5]. Meanwhile, whereas the real-space counterpart that we present in Eq. S4 remains underreported, it provides a powerful means to predict images recorded by scanning near-field optical microscopy.

We now briefly describe our method for evaluating $\hat{R}\Phi_{\rm dp}$ in the case of a spatially inhomogeneous 2D material at $z = 0$ described by a (piecewise) optical conductivity $\sigma_{2D}(\boldsymbol{\rho})$ upon a substrate with isotropic reflectivity $\beta_{\rm subs}$. We first consider the integro-differential equation for the scalar potential $\Phi_{\rm ref}$ generated by $\sigma_{2D}$ in response to the potential $\Phi_{\rm dp}$ of our quasi-dipolar probe[6], in absence of a substrate:

$$\left[1 + V * \frac{1}{2\pi q_p} \nabla \cdot \bar{\sigma}(\boldsymbol{\rho})\, \nabla\right] \Phi(\boldsymbol{\rho}) = \Phi_{\rm dp}(\boldsymbol{\rho}), \qquad \text{with } \Phi = \Phi_{\rm dp} + \Phi_{\rm ref}\,. \qquad \text{(S5)}$$

Here $q_p$ denotes the complex plasmon wavevector associated with our plasmonic medium and $\bar{\sigma}(\boldsymbol{\rho})$ is a piecewise homogeneous function equal to zero or 1 marking the lateral region $\boldsymbol{\rho} \in \Omega_p$ occupied by the plasmonic domain. Although in the present work we consider only a single-domain plasmonic medium, this prescription more generally allows to generate training data even for multi-domain plasmonic structures. Meanwhile, $V(\boldsymbol{r}, \boldsymbol{r}') = 1/|\boldsymbol{r} - \boldsymbol{r}'|$ is the Coulomb kernel, and the asterisk ($*$) denotes spatial convolution over the in-plane coordinate $\boldsymbol{\rho} = (x, y)$. We solve Eq. S5 by expanding $\Phi_{\rm ref}(\boldsymbol{\rho}) = \sum_n \phi_n^{\rm ref} \Phi_n(\boldsymbol{\rho})$ into an orthonormal basis of eigenfunctions specified on the domains $\Omega_p$ by $\nabla \cdot \bar{\sigma}(\boldsymbol{\rho})\, \nabla \Phi_n(\boldsymbol{\rho}) = -q_n^2 \Phi_n(\boldsymbol{\rho})$ and subject to the "zero current" boundary conditions $\hat{n} \cdot \nabla \Phi_n$ on the domain edges $\partial\Omega_p$, with $\hat{n}$ the unit vector normal to the domain's lateral boundary. These functions can be obtained with the finite element solver FEniCs[7] after meshing the experimentally relevant domain.

Values for the plasmon wave-vector are selected from the desired plasmon wavelength $\lambda_p$ and quality factor $Q$ according to $q_p = 2\pi(1 + iQ^{-1})/\lambda_p$.

Assembling the expansion coefficients $\phi_n^{\rm ref}$ into a vector $\boldsymbol{\phi}_{\rm ref}$, we solve Eq. (S5) by the matrix equation:

$$\boldsymbol{\phi}_{\rm ref} = -\left[\frac{-\boldsymbol{V}\boldsymbol{q}^2/(2\pi q_p)}{1-\boldsymbol{V}\boldsymbol{q}^2/(2\pi q_p)}\right] \boldsymbol{\phi}_{dp} \qquad \text{(S6)}$$

As with $\boldsymbol{\phi}_{\rm ref}$, here $\boldsymbol{\phi}_{dp}$ represents the vector of expansion coefficients for $\Phi_{\rm dp}(\boldsymbol{\rho})$. Meanwhile, each $\boldsymbol{q}^2$ denotes a diagonal matrix of eigenvalues $q_n^2$, and $\boldsymbol{V}$ is the coulomb matrix whose elements are given by $V_{m,n} = \int_{z=0} dA\, \Phi_m(\boldsymbol{\rho})\, V * \Phi_n(\boldsymbol{\rho})$. The term in brackets in Eq. (S6) represents the generalized reflection operator $\boldsymbol{R}$ for the system in the $\Phi_n$ basis. (The denominator is understood in the sense of a matrix inverse applied before pre-multiplication by the numerator.) While we defer the derivation to forthcoming work, this reflection operator generalizes to the case of a 2D material upon a substrate with isotropic reflectivity $\beta_{\rm subs}$ as follows:

$$\boldsymbol{R} = \frac{\beta_{\rm subs} - \boldsymbol{V}\boldsymbol{q}^2/(2\pi\kappa q_p)}{1-\boldsymbol{V}\boldsymbol{q}^2/(2\pi\kappa q_p)} \qquad \text{(S7)}$$

Here $\kappa = (\varepsilon_{\rm subs} + 1)/2$ and $\beta_{\rm subs} = (\varepsilon_{\rm subs} - 1)/(\varepsilon_{\rm subs} + 1)$, with $\varepsilon_{\rm subs}$ the substrate permittivity; in this work we take the value of 2.

We also define a symmetric matrix $\boldsymbol{Q}$ in the $\Phi_n$ basis corresponding to the spatial convolution in Eq. S4, with elements given by $Q_{m,n} = \int_{z=0} dA\, \Phi_m(\boldsymbol{\rho})\, q * \Phi_n(\boldsymbol{\rho})$. Since the $\Phi_n$ are orthonormal, Eq. S4 reduces to:

$$S(\boldsymbol{\rho}_{\rm dp}) \sim \frac{1}{2\pi} \boldsymbol{\phi}_{dp}(\boldsymbol{\rho}_{\rm dp})^T \, \boldsymbol{QR} \, \boldsymbol{\phi}_{dp}(\boldsymbol{\rho}_{\rm dp}). \qquad \text{(S8)}$$

This represents a vector norm of $\boldsymbol{\phi}_{dp}$ with respect to the matrix $\boldsymbol{QR}$.

In summary, after computing eigenfunctions $\Phi_n$ associated with plasmonic domain geometry, we compute symmetric matrices $\boldsymbol{V}$ and $\boldsymbol{Q}$ and the generalized reflectance operator $\boldsymbol{R}$. Then, in order to predict a spatial map $S(\boldsymbol{\rho}_{\rm dp})$, we simply project the incident potential emitted by our quasi-dipolar probe at each location $\boldsymbol{\rho}_{\rm dp}$ into the $\Phi_n$ basis by evaluating the vector of coefficients $\phi_{dp,n}(\boldsymbol{\rho}_{\rm dp}) = \int dA \, \Phi_{\rm dp}(\boldsymbol{\rho})\Phi_n(\boldsymbol{\rho})$ and successively applying Eq. S8. Although the eigenbasis $\Phi_n$ is of infinite size, projections into $\boldsymbol{\phi}_{dp}$ decay exponentially with $n$ when $\Phi_n$ are sorted by increasing eigenvalue $q_n^2$, so a truncated basis of size $N \approx 10^3$ is in our case sufficient for a converged map of near-field scattering amplitude $|S(\boldsymbol{\rho}_{\rm dp})|$. In this way, the observables of near-field microscopy can be predicted entirely by evaluating functions in the plane of the sample ($z = 0$). This computational method may be suitable for qualitative and quantitative modeling of near-field response of other spatially inhomogeneous 2D heterostructures. Such applications and details of their unique numerical implementation will be reported elsewhere.

## Architectures and training of convolutional neural network

| Layer(type) | Output Shape | Activation | Parameter |
|---|---|---|---|
| Conv2D | (398,398,32) | ReLU | 896 |
| Max_pooling2D | (199,199,32) | - | 0 |
| Conv2D | (197,197,64) | ReLU | 18496 |
| Max_pooling2D | (98,98,64) | Max pooling | 0 |
| Conv2D | (96,96,64) | ReLU | 36928 |
| Max_pooling2D | (48,48,64) | Max pooling | 0 |
| Conv2D | (46,46,128) | ReLU | 73856 |
| Max_pooling2D | (23,23,128) | Max pooling | 0 |
| Conv2D | (21,21,128) | ReLU | 147584 |
| Max_pooling2D | (10,10,128) | Max pooling | 0 |
| Conv2D | (8,8,128) | ReLU | 147584 |
| Max_pooling2D | (4,4,128) | Max pooling | 0 |
| Flatten | (2048) | - | 0 |
| Dropout(0.5) | (2048) | - | 0 |
| Dense | (256) | ReLU | 524544 |
| Dense | (128) | ReLU | 32896 |
| Dense | (2) | Linear | 1026 |

**Table S1. Architecture of the Convolutional neural network for polaritonic image analysis.**

Here, we describe the architecture of the convolutional neural network that is applied in this paper. The CNN is composed of feature extraction part and regression part, which are separated by the flatten layer. The feature extraction part of the CNN is a stack of convolution layers and max-pooling layers. In each convolution layer, convolutions (kernels) are operated on local patches (3×3) in input tensors and lead to an output 3D tensor, namely feature maps. The size of the feature maps is halved by a max-pooling layer, which effectively downsamples the feature maps and decreases the free parameters in the CNN. Since the kernels are much smaller than the input, the sparse weights additionally diminish the number of the free parameters in the CNN. The activation function that we adopted in all the convolution layers and the first two dense layers is the Rectified linear unit (ReLU) which endows the nonlinearity of the CNN. The activation function for the last dense layer is linear since the CNN aims at regression task. In the training process we implement the RMSprop algorithm as the optimizer and the learning rate is set to be 0.001. The batch size is set to be 50 and the Dropout layer with coefficient 0.5 is added to mediate the overfitting.

## Data augmentation and Data randomization

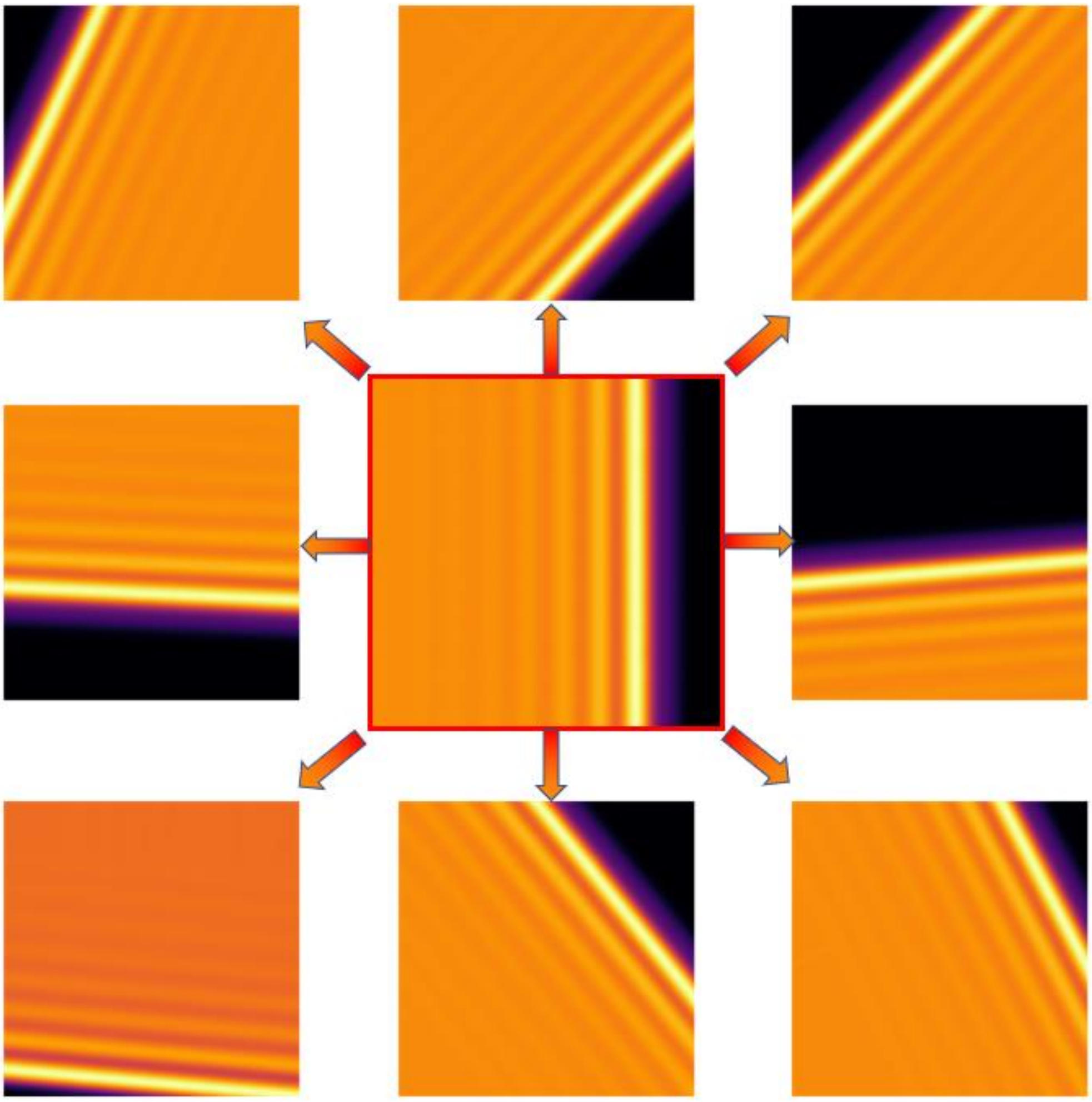

**Figure S1. Schematics of data augmentation. The volume of training dataset is enlarged by rotating and shifting the original simulation near-field image.**

The ability of the machine learning model is positively correlated to the volume of training dataset. In this paper, we apply the data augmentation on the training dataset. The near-field image was randomly rotated and shifted before being fed into the network in each epoch. The shift range is set to be 30% of the total width or height and the rotation angle is from 0 to 180 degrees. The newly created pixels after rotation and shifting are filled with colors of nearest pixels. After data augmentation, the training dataset is enlarged and the CNN is able to extract the parameters for near-field images with various sample edge positions and orientations. This data augmentation is crucial for the application since there is no need for data preprocessing before parameter extraction. To mediate the bias between experimental data and simulated data, noise and background signal are added, as shown in Figure S2. The substrate signal on the substrate is to imitate the complex environment outside the sample, which is not included in the simulation data. The amplitude of background signal and noise is evenly distributed in the range from 0% to 80% minimum near-field signal on the sample in original data.

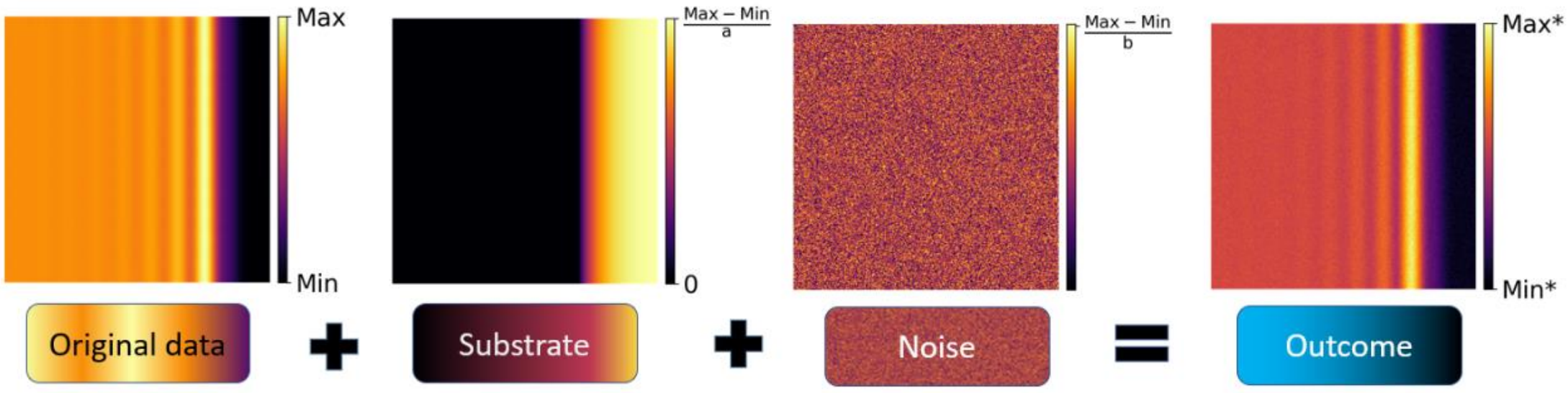

**Figure S2. Training data preprocessing. Substrate signal is added on the substrate area and noise is added on the whole area in the near-field image.**

## Experimental temperature dependent near-field images

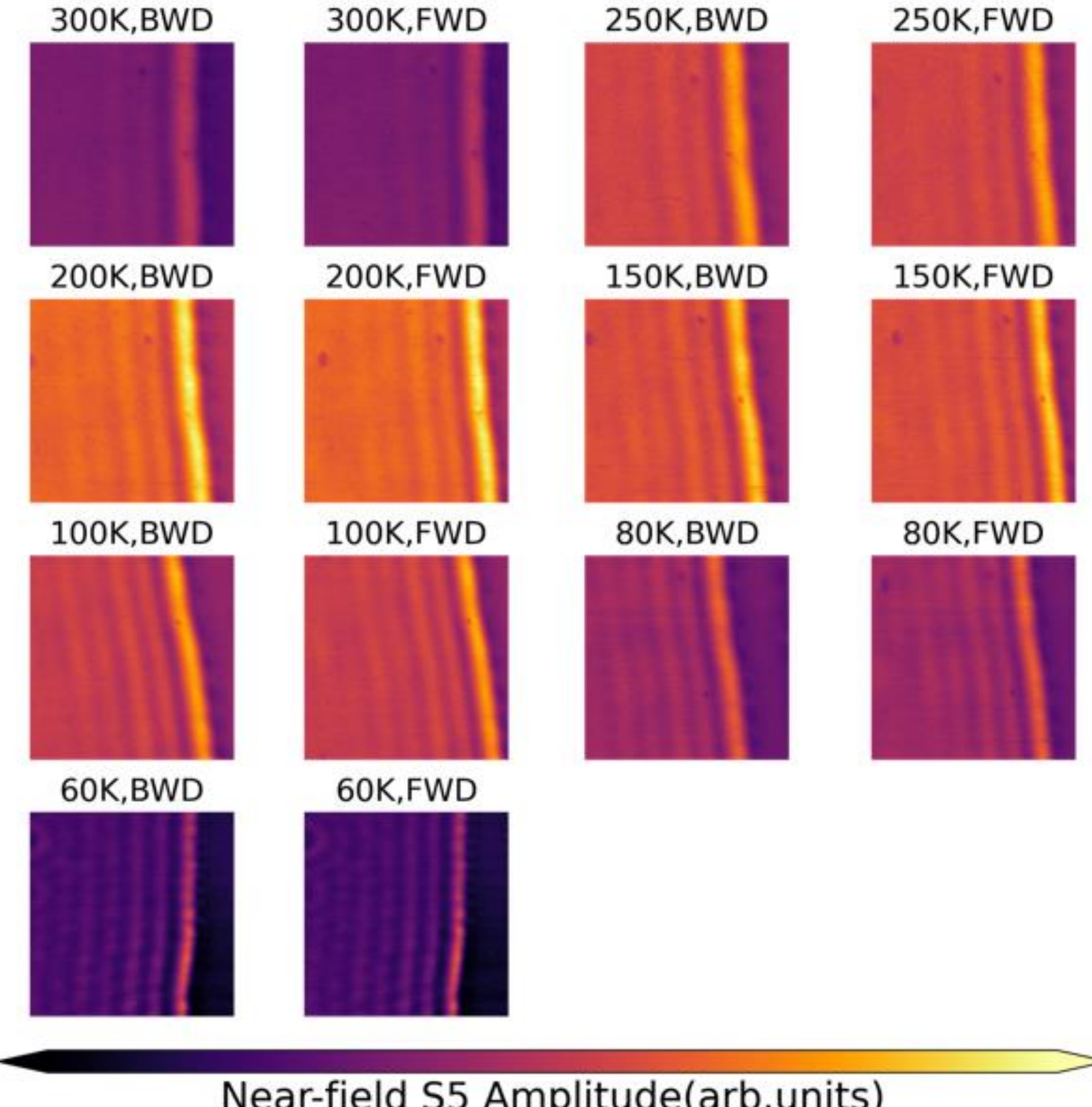

**Figure S3. Temperature dependent backward direction (BWD) scanned and forward direction (FWD) scanned near-field signal on hBN/Graphene/α-$RuCl_3$/$SiO_2$ heterostructure. The scales of pixels are slightly different between images measured at different temperatures and the final extracted wavelengths are rescaled corresponding to the pixel sizes.**

## Error distribution in parameter space

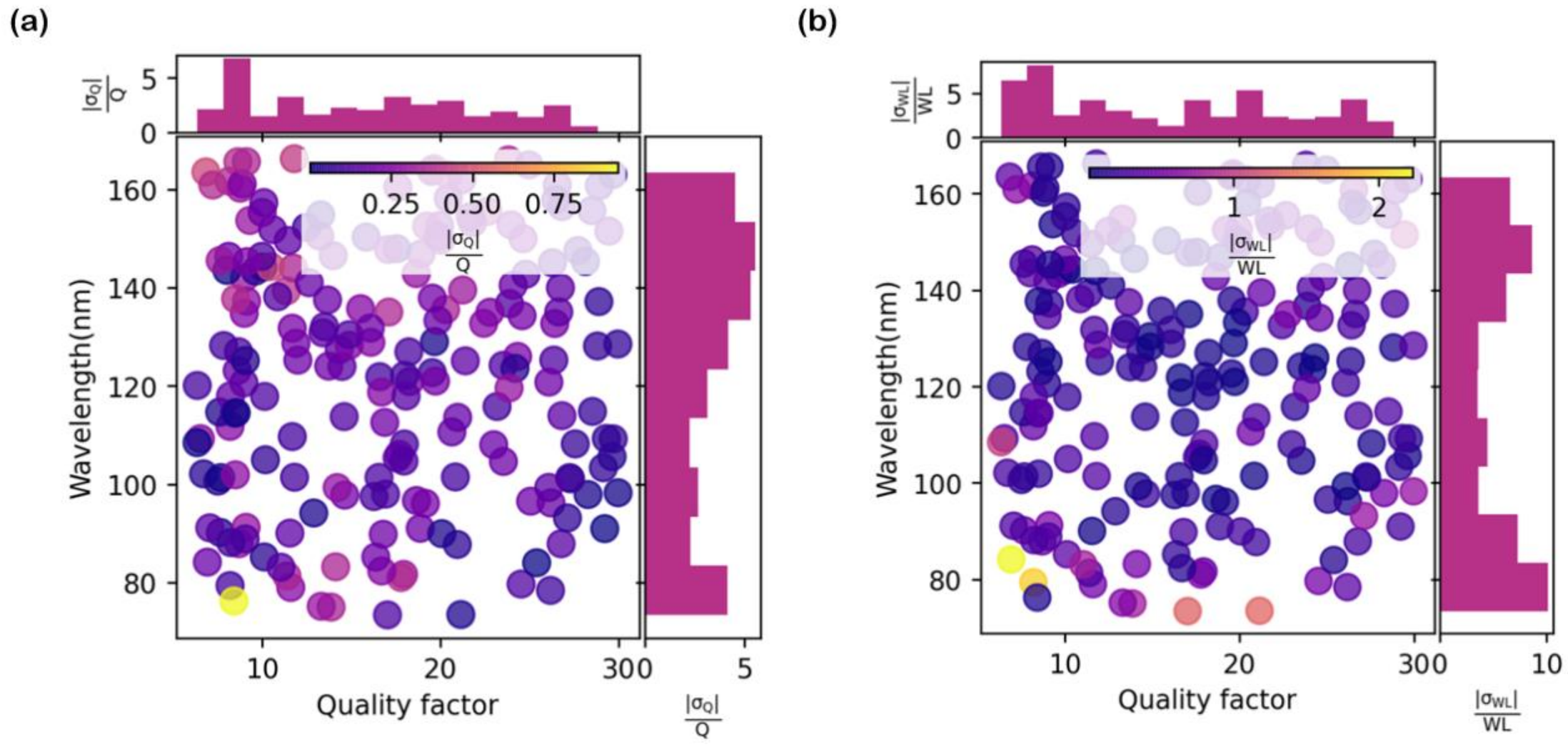

**Figure S4. Normalized absolute error distribution of (a) quality factor and (b) wavelength in parameter space.**

Here, we also investigate the error distribution of the current CNN in the parameter space. The error distribution is evaluated on the test dataset generated based on the same simulation method as the training data and the parameters are randomly selected and evenly distributed. As shown in figure S4(a) and (b), the extracted quality factor and wavelength tend to be inaccurate for large wavelength and small quality factor, respectively. The parameter dependent inaccuracy can be explained by fewer different plasmonic features in the confined simulation space when quality factor is small or the wavelength is large. A larger simulation area with perfect match boundary condition should be implemented in future data generation method. Since the test data is also synthetic, the bias between simulation and experiment still remains the same and this error distribution can't sufficiently evaluate the ability of the CNN to extract parameters in real experimental near-field images. A validation dataset composed of experimental data covered larger area in parameter space needs to be established as a standard test dataset for later network evaluation.

## Intermediate activation visualization

Unlike other machine learning models which behave like black-boxes and do not provide much readable information about the learning process, the CNN can present the extraction process by intermediate activation visualizations. Since the convolutional layers operate on feature maps, in which each channel can be viewed as a 2D map, the outcomes of convolutional layers can also be visualized as images. For example, the first convolutional layer sometimes acts as an edge detector. It was also demonstrated that the intermediate activations reflect the correlation function of the spin configuration in antiferromagnetic phase in the CNN for 2D Fermi-Hubbard model.[8] In figure S5, we present the intermediate activation visualization from the first convolutional layer, in which the sample and the substrate are separated and the profile of polaritonic fringes are sharpen. In deeper layers, the patterns presented in the intermediate activation visualization become more abstract but encode the information much closer to the final parameters, as shown in Figure S6. We also noticed that a lot of inactive channels appear in the current intermediate activation visualization, which indicates that the capability of the current network is not fully exploited and much more complex sample geometry can be involved in the training dataset.

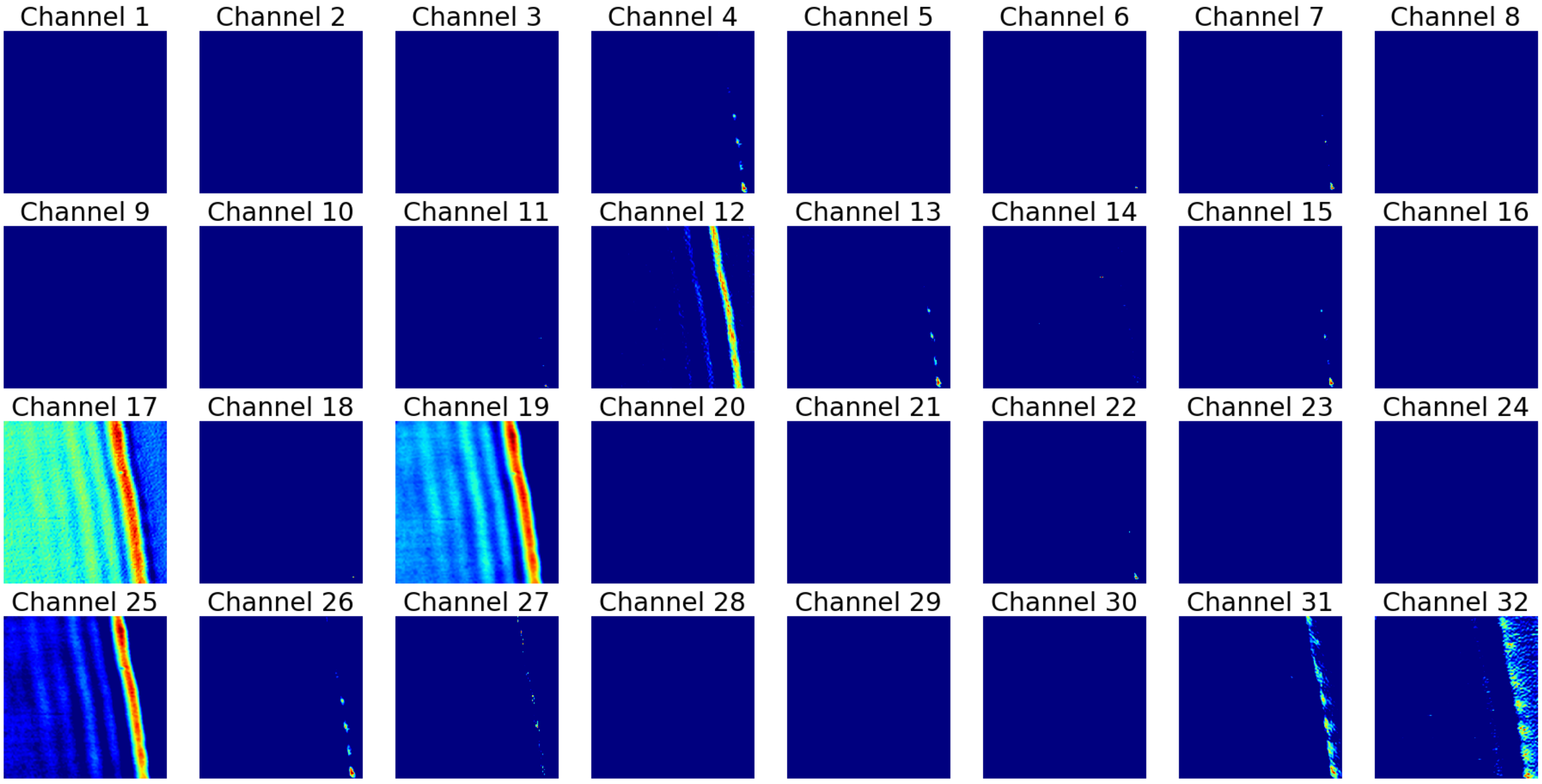

**Figure S5. Intermediate activation visualization from convolutional layer 1.**

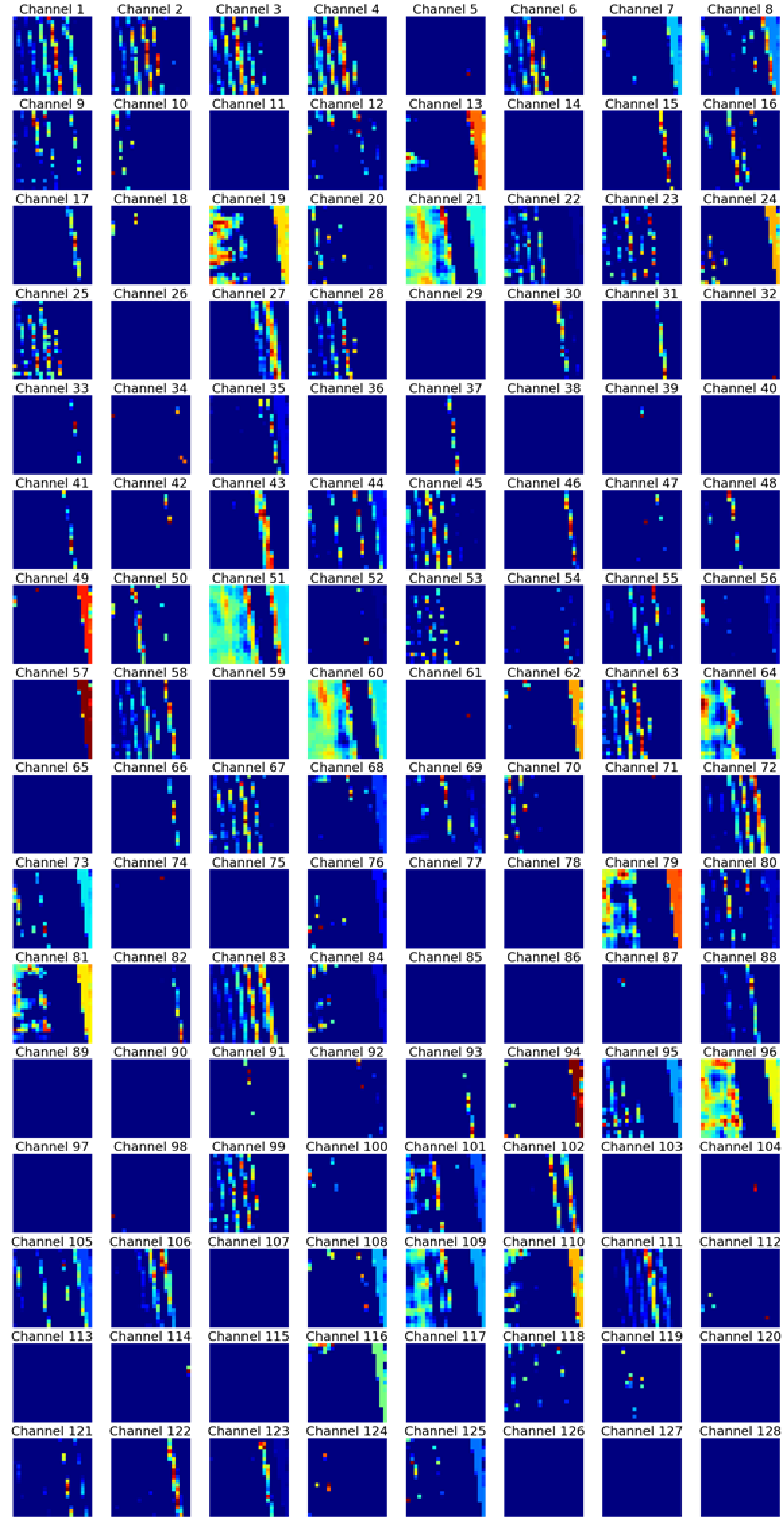

**Figure S6. Intermediate activation visualization from convolutional layer 5.**